\documentclass[11pt]{article}
\usepackage{hyperref}
\usepackage{amssymb}
\usepackage{graphicx}
\usepackage{slashed}
\usepackage{amsmath}
\usepackage{breqn}
\usepackage{authblk}
\usepackage{dsfont}
\usepackage{cite}
\usepackage[section]{placeins}
\usepackage[a4paper, total={6.8in, 10in}]{geometry}
\numberwithin{equation}{section}
\title{Quantisation of $\kappa$-deformed Dirac equation}
\author {E. Harikumar \thanks{harisp.uoh@nic.in} and Vishnu Rajagopal \thanks{vishnurajagopal.anayath@gmail.com}}
\affil{School of Physics, University of Hyderabad, \\Central University P.O, Hyderabad-500046, Telangana, India}
\date{}
\begin{document}

\maketitle
\begin{abstract}
In this paper, we study the quantisation of Dirac field theory in the $\kappa$-deformed space-time. We adopt a quantisation method that uses only equations of motion for quantising the field. Starting from $\kappa$-deformed Dirac equation, valid up to first order in the deformation parameter $a$, we derive deformed unequal time anti-commutation relation between deformed field and its adjoint, leading to undeformed oscillator algebra. Exploiting the freedom of imposing a deformed unequal time anti-commutation relations between $\kappa$-deformed spinor and its adjoint, we also derive a deformed oscillator algebra. We show that deformed number operator is the conserved charge corresponding to global phase transformation symmetry. We construct the $\kappa$-deformed conserved currents, valid up to first order in $a$, corresponding to parity and time-reversal symmetries of $\kappa$-deformed Dirac equation also. We show that these conserved currents and charges have a mass-dependent correction, valid up to first order in $a$. This novel feature is expected to have experimental significance in particle physics. We also show that it is not possible to construct a conserved current associated with charge conjugation, showing that the Dirac particle and its anti-particle satisfy different equations in $\kappa$-space-time.
\end{abstract}
\section{Introduction}
The notion of non-commutative space-time serves an elegant way of studying the gravitational effects at Planck length scales \cite{connes}. Different approaches to quantum gravity introduce minimal length scale, and non-commutativity provides a natural paradigm to incorporate such a length scale. Certain type of non-commutative space-time is shown to appear naturally in the low energy limit of string theory in \cite{sieberg}. Similarly, loop quantum gravity predicts the appearance of non-commutative space-time in its low energy limit. Quantum field theories in non-commutative space-times have been studied as a possible approach to handle the divergences \cite{snyder}. The field theories defined on non-commutative space-time is known to exhibit the mixing of UV and IR divergences \cite{Minwalla} and are known to be inherently non-local and non-linear. The underlying symmetry of the field theories on non-commutative space-time is known to be realised through Hopf algebra \cite{hopf}.

The two well studied non-commutative space-times are Moyal space-time \cite{sieberg} and $\kappa$-deformed space-times. The commutation relation between the space-time coordinates of Moyal space-time is a constant tensor, i.e.,

\begin{equation*}
 [\hat{x}_{\mu},\hat{x}_{\nu}]=\Theta_{\mu\nu}
\end{equation*}  
whereas the space-time coordinates of $\kappa$-deformed space-time obeys a Lie-algebra type commutation relation, i.e
\begin{equation*}
 [\hat{x}_{\mu},\hat{x}_{\nu}]=a_{\mu}\hat{x}_{\nu}-a_{\nu}\hat{x}_{\mu}.
\end{equation*}   
This space-time is shown to be the underlying space-time of deformed/doubly special relativity (DSR) \cite{majid}. DSR is an extension of the special theory of relativity, accomodating a fundamental length scale in a frame independent way \cite{kow}. This modification of the special theory of relativity was motivated by the fact that many different approaches to the theory of microscopic gravity predict the existence of a fundamental length scale. DSR is also known to modify the energy-momentum dispersion relation. It has been shown in \cite{wjk}, that the symmetry algebra of the $\kappa$-deformed space-time is defined by the $\kappa$-Poincare algebra, which is a Hopf algebra. The symmetry algebra of the $\kappa$-deformed space-time can also be realised using undeformed $\kappa$-Poincare algebra, where the defining relations are same as that of the usual Poincare algebra, but then the explicit form of the generators are deformed \cite{kreal,mel1,mel2,mel3,mel4}.

Various studies have been carried out analysing field theory models defined in $\kappa$-deformed space-time 
\cite{luk,mel5,mel6,pach,mel7,trg1,trg,hari,juric,hjm,sivakumar}. In most of these studies, field equations, which are invariant under the symmetry algebra, defined in $\kappa$-deformed space-times are set up from the $\kappa$-deformed quadratic Casimir \cite{kreal} of the algebra. Since the quadratic Casimir has higher-order derivatives, the Lagrangian corresponding to these deformed field equations also contains higher-order derivative terms in it, signalling the non-local nature of the non-commutative field theories and make canonical quantisation difficult.

The knowledge of the explicit form of the Lagrangian is necessary for applying the canonical scheme to quantise field theory, but there exists an alternate quantisation procedure which does not require the Lagrangian for quantising field theories. Instead, this method uses the equations of motion of the field theory as the starting point for the quantisation \cite{yt,ume,ytaka}. The quantisation of massive spin-one field has been studied using this method \cite{pal}. In \cite{vis} covariant commutation relation for field describing an arbitrary spin had been obtained using this method. In this method, every free field equation of motion is converted to Klein-Gordon equation with the help of an operator called Klein-Gordon divisor \cite{yt,ume,ytaka}. This Klein-Gordon divisor is then used to define an unequal time anti-commutation relation between the spinor field operator and its adjoint operator such that the field equations are consistent with the Heisenberg's equations of motion and it leads to the usual (anti-) commutation relations between the creation and annihilation operators, which appears in the decomposition of the field operator. In this procedure, the equations of motion facilitate the derivation of the conserved currents corresponding to the symmetry transformation, without referring Lagrangian \cite{ytaka,lur}. One of the most interesting features of this method is that it provides a way to construct conserved currents corresponding to the discrete symmetries \cite{lur}, which is not possible using Noether's prescription. 

The Lagrangian corresponding to non-commutative field theories involves higher-order derivatives, and therefore it is 
difficult to quantise them using the conventional canonical quantisation method. Also, the equations of motion 
corresponding to different field theories defined in $\kappa$-deformed space-time have been constructed without referring 
to their Lagrangians. The $\kappa$-deformed Maxwell's equations have been derived in \cite{hari}, using Feynman's 
approach and in \cite{juric} Maxwell's equations in $\kappa$-deformed space-time have been derived by elevating the 
principle of minimal coupling to $\kappa$-deformed space-time. In \cite{hjm} the $\kappa$-deformed geodesic equation 
has been derived by generalising the Feynman's approach to $\kappa$-deformed space-time. In \cite{sivakumar} the 
$\kappa$-deformed Dirac equation has been derived by replacing the commutative derivative with Dirac derivative, 
which transform as a four-vector under undeformed $\kappa$-Poincare algebra. In all these works, the equations of 
motion were obtained, but not the Lagrangian. Different gauge theory models in $\kappa$-deformed space-time has been 
constructed and studied in \cite{meyer}. In \cite{vishnu1} the quantisation of the $\kappa$-deformed Klein-Gordon 
field has been studied using the procedure mentioned above, resulting in a $\kappa$-deformed oscillator algebra, which modifies the number operator and thus modifying the Unruh effect also. Thus it is of intrinsic interest to 
study quantisation whose starting point is the equation of motion and not the Lagrangian.

 Dirac Equation in the $\kappa$-space-time has been constructed and studied by various authors \cite{wjk,luk, now,dirac-mom,diracap, diraco,ravi2}, using different approaches. The $\kappa$-deformed Dirac equation constructed in  \cite{wjk,luk}, did give $\kappa$-deformed Klein-Gordon equation on squaring as in the commutative space-time but was not invariant under spin-$\frac{1}{2}$ representation of $\kappa$-Poincare algebra.  A $\kappa$-deformed Dirac equation which is invariant under $\frac{1}{2}$-spin representation of $\kappa$-Poincare algebra was constructed in  \cite{now}, which was related to Pauli-Lubanski four-vector rather than the energy-momentum Casimir of the deformed algebra (which is square of  Dirac equation in the commutative space-time). In \cite{dirac-mom}, Dirac equation in momentum space,  compatible with deformed special relativity was constructed and since $\kappa$-deformed space-time was the underlying space-time of deformed special relativity, this construction gave $\kappa$-deformed Dirac equation in momentum space. Another equation describing spin $\frac{1}{2}$ particles compatible with deformed special relativity was proposed in \cite{diracap}, but the corresponding anti-particle equation has shown to be different from this. In \cite{sivakumar}, A Dirac equation consistent with undeformed $\kappa$-Poincare algebra was obtained, and its implications were analysed. It was shown that the charge conjugation is not a symmetry of this equation. The quantisation of this Dirac field, following the method used for quantisation of $\kappa$-deformed Klein-Gordon field used in \cite{trg1, trg} was carried out in \cite{ravi2}, showing that the creation and annihilation operators satisfy a deformed algebra. This was derived by demanding the compatibility between the twisted flip operator and the Hopf algebra structure of the undeformed $\kappa$-Poincare algebra. All these deformed Dirac equations \cite{wjk,luk, now,dirac-mom,diracap, diraco,ravi2}, in the commutative limit, reproduced the expected results. Thus it is interesting and important to study the quantisation of deformed Dirac equation using different approaches and analyses them. It is also important to note that Lagrangians describing $\kappa$-deformed field theory models are not uniquely defined. Reduction of deformed Lagrangian to correct commutative limit is used as a guiding principle, and this leaves room for non-unique results, and thus it is intrinsically important to study the quantisation using field equations, rather than Lagrangian (or action).

In this paper, we use the above-described procedure \cite{yt,ume,ytaka} for quantising the $\kappa$-deformed Dirac field, starting from its field equation, valid up to first order in $a$. This $\kappa$-deformed equation was constructed by replacing, in the usual Dirac equation, the commutative derivative with Dirac derivative, which transforms as a four-vector under undeformed $\kappa$-Poincare algebra. We derive the deformed unequal time anti-commutation relation between the deformed field operator and its adjoint, using the usual oscillator algebra. We further obtain a deformed oscillator algebra which keeps the unequal time anti-commutation relation between the $\kappa$-deformed field and it is adjoint undeformed.  This deformed oscillator algebra is a novel feature, and it comes out naturally as in $\kappa$-deformed Klein-Gordon field \cite{vishnu1}.  We show that the deformed number operator is the conserved charge corresponding to the global phase transformation symmetry and also derive the deformed conserved currents corresponding to the parity and the time-reversal symmetry of the $\kappa$-deformed Dirac equation, valid up to first order in $a$. We show that the correction to these conserved charges due to $\kappa$-deformation have one part which also depends on the mass of the particle. This novel feature is of interest as it can lead to interesting consequences in particle physics. We also show that the charge conjugation is not a symmetry of the Dirac equation in $\kappa$-space-time. 

The organisation of the paper is as follows. In sec. 2, we briefly discuss the quantisation procedure that uses only the equations of motion as the starting point and defines anti-commutation relation between the spinor field and its adjoint in terms of the Klein-Gordon divisor, which acting on the equation of motion, transforms it into Klein-Gordon equation. Using this equation of motion, the expression for the conserved current, corresponding to a symmetry transformation, is then defined. In sec. 3 we summarise the construction of the $\kappa$-deformed Dirac equation, valid up to first order in $a$, by replacing commutative derivative with Dirac derivative, such that product of deformed field equation and its adjoint equation gives the $\kappa$-deformed Klein-Gordon equation, valid up to first order in $a$. We also obtain the $\kappa$-deformed Klein-Gordon divisor. In sec. 4 we solve the $\kappa$-deformed Dirac equation perturbatively, valid up to first order in $a$, and using this we derive the deformed unequal time anti-commutation relation between deformed field and its adjoint, using the undeformed oscillator algebra. We also derive the deformed oscillator algebra by demanding that the unequal time anti-commutation relation between deformed field and its adjoint is unaffected by the $\kappa$-deformation of space-time. In sec. 5 we show that the deformed number operator is the conserved charge corresponding to global phase transformation symmetry of a deformed Dirac field. In subsec. 5.1 and 5.2, we derive the deformed conserved currents corresponding to parity and time-reversal symmetry of $\kappa$-deformed Dirac equation, respectively.  We show here that one of the correction terms depends on the mass of the Dirac particle, which is a novel result. In subsec. 5.3 we show that charge conjugation is not a symmetry of $\kappa$-deformed Dirac equation, valid up to first order in $a$. In sec. 6, we discuss our results and present conclusions. In appendix A, we show that in the limit of commutative space-time the conserved charges obtained above reduce to the generators of the corresponding symmetry transformations. In appendix B, we derive the $\kappa$-deformed energy-momentum tensor and using it obtain the deformed conserved four momenta and the angular momenta, valid up to first order in $a$.

Here we use $\eta_{\mu\nu}=diag(-1,1,1,1)$.

\section{Quantisation without Lagrangian}

In this section, we give a brief review of the quantisation procedure introduced by Y. Takahashi and H. Umezawa in \cite{yt}. This method does not require the explicit form of the Lagrangian for quantisation of the fields but needs only the equations of motion satisfied by the field for its quantisation.

The equations of motion corresponding to a free field and its adjoint can be expressed as
\begin{equation}\label{a1}
 \Lambda(\partial)\phi(x)=0,~~\bar{\phi}(x)\Lambda(-\overleftarrow{\partial})=0, 
\end{equation}
respectively. Here $\Lambda(\partial)$ is a polynomial of derivative operators $\partial_{\mu}$, whose explicit form is given as
\begin{equation}
 \begin{split}
\Lambda(\partial)&=\sum_{l=0}^{N}{\Lambda_{\mu_1\mu_2....\mu_l}}\partial_{\mu_1}\partial^{\mu_2}....\partial^{\mu_l}\\
&=\Lambda+\Lambda_{\mu}\partial^{\mu}+\Lambda_{\mu\nu}\partial^{\mu}\partial^{\nu}+\Lambda_{\mu\nu\rho}\partial^{\mu}\partial^{\nu}\partial^{\rho}+.....+\Lambda_{\mu_1\mu_2\mu_3.....\mu_N}\partial^{\mu_1}\partial^{\mu_2}\partial^{\mu_3}.........\partial^{\mu_N}
\end{split}
\end{equation}
The above construction is valid when the $\Lambda$'s are symmetric in its indices. 
 
According to this procedure every free field equations of motion reduces to Klein-Gordon equation when it is acted upon by an operator called Klein-Gordon divisor, denoted by $d(\partial)$, such that it satisfy the relations
\begin{equation}\label{a02}
\Lambda(\partial)d(\partial)=\Box-m^2,  ~~[\Lambda(\partial),d(\partial)]=0, ~~det[d(\partial)]\neq 0.
\end{equation}
For Klein-Gordon field, $d(\partial)$ is the identity operator-${\mathds 1}$. Similarly for a Dirac field, we have $\Lambda(\partial)=i\slashed{\partial}+m$ and $d(\partial)=i\slashed{\partial}-m$.

The field operator satisfying the equation of motion given in Eq.(\ref{a1}) is decomposed using the creation and the annihilation operator as  
\begin{equation}\label{a3}
\phi(x)=\int\frac{d^3p}{\sqrt{(2\pi)^32E_p}}\Big(u_p(x)a(p)+\bar{u}_p(x)a^{\dagger}(p)\Big).
\end{equation}
Here, $u_p(x)$ satifies $ \Lambda(\partial)u_p(x)=0$. For a bosonic field, the creation and annihilation operators appearing in Eq.(\ref{a3}) satisfy canonical commutation relation. Similarly for a fermionic field they obey anti-commutation relations given by
\begin{equation}\label{a4}
[a(k),a(k')]_+=[a^{\dagger}(k),a^{\dagger}(k')]_+=0,~~ [a(k),a^{\dagger}(k')]_+=\delta^3(k-k').
\end{equation}
The field operator, defined in Eq.(\ref{a3}),  and its adjoint operator follow an unequal time (anti-)commutation relation in the light cone. This canonical (anti-)commutation relation is compatible with Heisenberg's equation of motion\cite{ytaka,yt}, 
\begin{equation}
 i\partial_{\mu}\phi(x)=[\phi(x),P_{\mu}]
\end{equation}
satisfied by the field operator $\phi(x)$. The unequal time anti-commutation relation (if the corresponding creation and annihilation operators follow Eq.(\ref{a4})) between spinor field and its conjugate is given by
\begin{equation}\label{a5}
[\phi(x),\bar{\phi}(x')]_+=id(\partial)\Delta(x-x')
\end{equation}
where 
\begin{equation*}
 \Delta(x-x')=\int\frac{d^3p}{(2\pi)^32E_p}\big(e^{-ip(x-x')}-e^{ip(x-x')}\big).
\end{equation*}
The above quantisation scheme also provides a unique way of constructing the conserved current from the corresponding equations of motion, without using its Lagrangian \cite{ytaka,yt,lur}. The conserved current is derived with the help an operator, $\Gamma_{\mu}(\partial,-\overleftarrow{\partial})$ which is defined as
\begin{equation}\label{a6}
\begin{split}
 \Gamma_{\mu}(\partial,-\overleftarrow{\partial})&=\sum_{l=0}^{N-1}\sum_{i=0}^{l}\Lambda_{\mu\mu_1.....\mu_l}\partial_{\mu_1}.....\partial_{\mu_i}(-\overleftarrow{\partial}_{\mu_{i+1}})......(-\overleftarrow{\partial}_{\mu_l})\\
&=\Lambda_{\mu}+\Lambda_{\mu\nu}(\partial^{\nu}-\overleftarrow{\partial}^{\nu})+\Lambda_{\mu\nu\rho}(\partial^{\nu}\partial^{\rho}-\partial^{\nu}\overleftarrow{\partial}^{\rho}+\overleftarrow{\partial}^{\nu}\overleftarrow{\partial}^{\rho})+......
\end{split}
\end{equation}
and it satisfies the identity
\begin{equation}\label{iden}
 (\partial^{\mu}+\overleftarrow{\partial}^{\mu})\Gamma_{\mu}(\partial,-\overleftarrow{\partial})=\Lambda(\partial)-\Lambda(-\overleftarrow{\partial}).
\end{equation}
This identity allows us to verify that the conserved current, corresponding to the symmetry of the field equation, is divergenceless. 

Let us assume that the field equation and its adjoint are invariant under the transformations given by
\begin{equation}
 \phi(x)\rightarrow F[x]\textnormal{ and }\bar{\phi}(x)\rightarrow G[x]
\end{equation} 
where $F[x]$ and $G[x]$ are functionals of the field operator. Under the above transformations, the equations of motion corresponding to field and its adjoint becomes
\begin{equation}
 \Lambda(\partial)F[x]=0 \textnormal{ and }G[x]\Lambda(-\overleftarrow{\partial})=0,
\end{equation}
respectively. Now using the operator $\Gamma_{\mu}(\partial,-\overleftarrow{\partial})$, defined in Eq.(\ref{a6}), the conserved current is defined as
\begin{equation}\label{conser}
 J_{\mu}(x)=G[x]\Gamma_{\mu}(\partial,-\overleftarrow{\partial})F[x].
\end{equation}
Thus constructed conserved current provides a definition of normalisation of field operators. Using this procedure one can also obtain the conserved currents corresponding to the discrete symmetries.
\section{$\kappa$-deformed Dirac equation}
In this section we summarise the $\kappa$-deformed Dirac equation, valid up to first order in $a$ \cite{sivakumar}. The $\kappa$-deformed space-time coordinates satisfy a Lie-algebra type commutation relations given by
\begin{equation}\label{nc}
[\hat{x}_i,\hat{x}_j]=0,~~[\hat{x}_0,\hat{x}_i]=ia\hat{x}_i
\end{equation}
where $a$ is the $\kappa$-deformation parameter having the dimension of length and we choose a specific realisation for the $\kappa$-deformed space-time \cite{kreal}, i.e.,
\begin{equation}\label{real}
 \hat{x}_{\mu}=x_{\alpha}\varphi^{\alpha}_{\mu}.
\end{equation}
Substituting Eq.(\ref{real}) in Eq.(\ref{nc}), we get realisation for $\kappa$-deformed space-time coordinate as 
\begin{equation}\label{soln}
 \varphi _0^0=1, \, \varphi _i^0=0, \, \varphi_0^i=0, \, \varphi _j^i=\delta _j^i e^{-A}, 
\end{equation}
where $A=ap_0$. The symmetry algebra of the $\kappa$-space-time is defined through undeformed $\kappa$-Poincare algebra \cite{kreal,mel1}. Since the commutative derivatives do not transform as $4$-vector under undeformed $\kappa$-Poincare algebra, Dirac derivatives are defined for this purpose and they are given by
\begin{equation}\label{dr1} 
D_{i}=\partial_{i}\frac{e^{-A}}{\varphi},~~
D_{0}=\partial_{0}\frac{\textnormal{sinh}A}{A}-ia\partial_{i} ^2 \frac{e^{-A}}{2\varphi^2}.
\end{equation}
The quadratic Casimir of the undeformed $\kappa$-Poincare algebra is constructed using Dirac derivatives as
\begin{equation}\label{cas}
D_{\mu}D^{\mu}= \Box\Big(1+\frac{a^2}{4}\Box\Big),
\end{equation}
where $\Box=\partial_{i} ^2 \frac{e^{-A}}{\varphi^2}+2\partial_{0} ^2\frac{(1-\textnormal{cosh}A)}{A^2}$. Thus we construct the $\kappa$-deformed Klein-Gordon equation using Eq.(\ref{cas}) as
\begin{equation}\label{kg}
\bigg(\Box\Big(1+\frac{a^2}{4}\Box\Big)-m^2\bigg)\hat{\phi}(x)=0.
\end{equation}
Up to first order in $a$, this $\kappa$-deformed Klein-Gordon equation becomes
\begin{equation}\label{kg1}
 \Big(\partial_i ^2-\partial_0 ^2-m^2-ia\partial_0\partial_i ^2\Big)\hat{\phi}(x)=0.
\end{equation}
Now we obtain the $\kappa$-deformed Dirac equation and its conjugate equation, by replacing the commutative derivative with Dirac derivative, as 
\begin{equation}\label{d1}
 \Big(iD_{\mu}\gamma^{\mu}+m\Big)\hat\psi(x)=0,
\end{equation}
and
\begin{equation}\label{d2}
 \bar{\hat\psi}(x)\Big(i\overleftarrow{D}_{\mu}\gamma^{\mu}-m\Big)=0.
\end{equation}
The product of Eq.(\ref{d1}) and Eq.(\ref{d2}) gives the $\kappa$-deformed Klein-Gordon equation, i.e., Eq.(\ref{kg}). Up to first order in $a$, Eq.(\ref{d1}) and Eq.(\ref{d2}) become
\begin{equation}\label{d3}
 \Big(i\partial_{\mu}\gamma^{\mu}+\frac{a}{2}\gamma^0\partial_i^2+m\Big)\hat\psi(x)=0,
\end{equation}
\begin{equation}\label{d4}
 \bar{\hat\psi}(x)\Big(i\overleftarrow{\partial}_{\mu}\gamma^{\mu}+\frac{a}{2}\gamma^0\overleftarrow{\partial}_i^2-m\Big)=0,
\end{equation}
and their product gives Eq.(\ref{kg1}) (here we have used $[\gamma^{\mu},\gamma^{\nu}]_+=-2\eta ^{\mu\nu}$), i.e.,
\begin{equation}\label{lamb1}
  \Big(i\partial_{\mu}\gamma^{\mu}+\frac{a}{2}\gamma^0\partial_i^2+m\Big)\Big(i{\partial}_{\nu}\gamma^{\nu}+\frac{a}{2}\gamma^0{\partial}_i^2-m\Big)=\Box-ia\partial_0\partial_i^2-m^2.
\end{equation}
In the $\kappa$-deformed space-time, valid up to first order in $a$, first of Eq.(\ref{a02}) becomes  
\begin{equation}\label{lamb}
 \hat{\Lambda}(\partial)\hat{d}(\partial)=\Box-m^2-ia\partial_0\partial_i ^2.
\end{equation}
Comparing Eq.(\ref{lamb1}) with Eq.(\ref{lamb}), we identify 
\begin{equation}\label{lambda}
 \hat{\Lambda}(\partial)=i{\partial}_{\nu}\gamma^{\nu}+\frac{a}{2}\gamma^0{\partial}_i^2+m,
\end{equation}
and $\kappa$-deformed Klein-Gordon divisor as 
\begin{equation}
 \hat{d}(\partial)=i\partial_{\mu}\gamma^{\mu}+\frac{a}{2}\gamma^0\partial_i^2-m.
\end{equation}
\section{Quantisation of $\kappa$-deformed spinor field}
In this section we quantise the Dirac equation defined in the $\kappa$-deformed space-time by applying the above procedure \cite{yt,ume,ytaka} and obtain the corresponding deformed oscillator algebra, valid up to first order in $a$. We also show that by requiring the unequal time anti-commutation relation between spinor and its conjugate to be undeformed, we obtain a deformation of the oscillator algebra. 

The $\kappa$-deformed field operator and its adjoint are decomposed using creation and anihilation operators as
\begin{equation}\label{decom1}
 \hat{\psi}(x)=\int \frac{d^3p}{\sqrt{(2\pi)^32E_p}}\sum_{s=1,2}\Big(\hat{a}_s(p)\hat{u}_s(p)e^{-ipx}+\hat{b}^{\dagger}_s(p)\hat{v}_s(p)e^{ipx}\Big)
\end{equation}
and
\begin{equation}\label{decom2}
 \bar{\hat{\psi}}(x)=\int \frac{d^3p}{\sqrt{(2\pi)^32E_p}}\sum_{s=1,2}\Big(\hat{a}^{\dagger}_s(p)\bar{\hat{u}}_s(p)e^{ipx}+\hat{b}_s(p)\bar{\hat{v}}_s(p)e^{-ipx}\Big),
\end{equation}
respectively. Here $E_p=\sqrt{p^2+m^2}$, is the commutative disperison relation. $\hat{\psi}(x)$ and $\bar{\hat{\psi}}(x)$ defined in Eq.(\ref{decom1}) and Eq.(\ref{decom2}) satisfy Eq.(\ref{d3}) and Eq.(\ref{d4}), respectively. We find that $\hat{u}_s(p)e^{-ipx}$ and $\hat{v}_s(p)e^{ipx}$ obey the $\kappa$-deformed Dirac equation, i.e., $\hat{\Lambda}(\partial)\big(\hat{u}_s(p)e^{-ipx}\big)=0$ and $\hat{\Lambda}(\partial)\big(\hat{v}_s(p)e^{ipx}\big)=0$. Now we solve these equations perturbatively by expanding $\hat{u}_s(p)$ and $\hat{v}_s(p)$ up to first order in $a$, as
\begin{equation}
 \hat{u}(p)=u^{(0)}(p)+a\alpha u^{(1)}(p),~~~\hat{v}(p)=v^{(0)}(p)+a\alpha v^{(1)}(p),
\end{equation}
where $\alpha$ is a parameter having the dimension of $[L]^{-1}$, i.e.,
\begin{equation}\label{d6}
 \Big(i\partial_{\mu}\gamma^{\mu}+\frac{a}{2}\gamma^0\partial_i^2+m\Big)\big(u^{(0)}(p)+a\alpha u^{(1)}(p)\big)e^{-ipx}=0,
\end{equation}
and
\begin{equation}\label{d7}
 \Big(i\partial_{\mu}\gamma^{\mu}+\frac{a}{2}\gamma^0\partial_i^2+m\Big)\big(v^{(0)}(p)+a\alpha v^{(1)}(p)\big)e^{ipx}=0.
\end{equation}
Let us consider Eq.(\ref{d6}) first. By separating the $a$ independent and $a$ dependent terms, we get two equations as
\begin{equation}\label{a2}
 \Big(i\gamma^{\mu}\partial_{\mu}+m\Big)u^{(0)}(p)e^{-ipx}=0;~~
 \Big(i\alpha\gamma^{\mu}\partial_{\mu}+\alpha m\Big)u^{(1)}(p)e^{-ipx}+\frac{1}{2}\gamma^0\partial_i^2u^{(0)}e^{-ipx}=0.
\end{equation}
After re-writing, second of Eq.(\ref{a2}) as
\begin{equation}
 \Big(i\gamma^{\mu}\partial_{\mu}+m\Big)u^{(1)}(p)e^{-ipx}=\frac{1}{2\alpha}\gamma^0p_i^2u^{(0)}e^{-ipx},
\end{equation}
we use Green's function method to solve this inhomogeneous differential equation and the solution is obtained as
\begin{equation}\label{c1}
 u^{(1)}e^{-ipx}=u^{(0)}e^{-ipx}+\int d^4x'G(x-x')j(x'),
\end{equation}
where, $\Big(i\gamma^{\mu}\partial_{\mu}+m\Big)G(x-x')=\delta^4(x-x')$ and $j(x)=\frac{1}{2\alpha}\gamma^0p_i^2u^{(0)}e^{-ipx}$. The Green's function $G(x-x')$ is given by
\begin{equation}
 G(x-x')=-\int \frac{d^4p}{(2\pi)^4}\frac{\slashed{p}-m}{p^2+m^2}e^{-ip(x-x')}.
\end{equation}
We shift the poles by $i\epsilon$ and evaluate the integral as 
\begin{equation}
\begin{split}
 G(x-x')=&-\int \frac{d^4p}{(2\pi)^4}\frac{\slashed{p}-m}{p^2+m^2+i\epsilon}e^{-ip(x-x')}\\
=&-\int \frac{d^4p}{(2\pi)^4}\frac{i\slashed{\partial}-m}{p^2+m^2+i\epsilon}e^{-ip(x-x')}\\
=&-(i\slashed{\partial}-m)\int \frac{d^4p}{(2\pi)^4}\frac{1}{p^2+m^2+i\epsilon}e^{-ip(x-x')}\\
=&(i\slashed{\partial}-m)\Delta_F(x-x')
\end{split}
\end{equation}
where $\Delta_F(x-x')=-\int \frac{d^4p}{(2\pi)^4}\frac{1}{p^2+m^2+i\epsilon}e^{-ip(x-x')}$ is the propagator for the complex Klein-Gordan equation and from \cite{vishnu1} we find 
\begin{equation}\label{c2}
 G(x-x')=-(i\slashed{\partial}-m)\int \frac{d^3p}{(2\pi)^3}\frac{i}{2E_p}\Big[\theta(t-t')e^{-ip(x-x')}-\theta(t'-t)e^{ip(x-x')}\Big].
\end{equation}
Substituting Eq.(\ref{c2}) in Eq.(\ref{c1}), we get
\begin{multline}
 u^{(1)}(p)e^{-ipx}=u^{(0)}(p)e^{-ipx}-i\int \frac{d^4x'}{(2\pi)^3}\int\frac{d^3p'}{2E_{p'}}\Big((i\slashed{\partial'}-m)\Big[\theta(t-t')e^{-ip'(x-x')}\\-\theta(t'-t)e^{ip'(x-x')}\Big]\Big)\frac{p_i^2}{2\alpha}\gamma^0u^{(0)}(p)e^{-ipx'}
\end{multline}
\begin{multline}
=u^{(0)}(p)e^{-ipx}-\frac{i}{2\alpha}\int\frac{d^4x'}{(2\pi)^3}\int\frac{d^3p'}{2E_{p'}}\Big[-i\delta(t-t')e^{-ip'(x-x')}-i\delta(t'-t)e^{ip'(x-x')}\Big]p_i^2u^{(0)}(p)e^{-ipx'}+\\\frac{i}{2\alpha}\int \frac{d^4x'}{(2\pi)^3}\int\frac{d^3p'}{2E_{p'}}\Big[(\slashed{p}'+m)\theta(t-t')e^{-ip'x}e^{-i(p-p')x'}+(\slashed{p}'-m)\theta(t'-t)e^{ip'x}e^{-i(p+p')x'}\Big]\gamma^0p_i^2u^{(0)}(p)
\end{multline}
\begin{multline}\label{new}
=u^{(0)}(p)e^{-ipx}-\frac{1}{2\alpha}\int\frac{d^3p'}{2E_{p'}}\Big[e^{-ip'\cdot x}e^{iE_pt}\int\frac{d^3x'}{(2\pi)^3}e^{i(p'-p)\cdot x'}+e^{ip'\cdot x}e^{iE_pt}\int\frac{d^3x'}{(2\pi)^3}e^{-i(p'+p)\cdot x'}\Big]p_i^2u^{(0)}(p)+\\\frac{i\pi}{\alpha}\int \frac{d^3p'}{2E_{p'}}\Big[(\slashed{p}'+m)\theta(t-t')e^{-ip'x}\int \frac{d^4x'}{(2\pi)^4}e^{-i(p-p')x'}+(\slashed{p}'-m)\theta(t'-t)e^{ip'x}\int \frac{d^4x'}{(2\pi)^4}e^{-i(p+p')x'}\Big]\gamma^0{p_i^2}u^{(0)}(p)
\end{multline}
Now we use integral representation of step function, i.e.,, $\Theta(t-t')=\lim_{\epsilon\to 0}\frac{1}{2\pi i}\int dk \frac{e^{ik(t-t')}}{k-i\epsilon}$ and the identity, $\delta(x-a)f(x)=\delta(x)f(a)$ in the last two terms on RHS of Eq.(\ref{new}) and we get
\begin{equation}\label{c2a}
u^{(1)}(p)e^{-ipx}=u^{(0)}(p)e^{-ipx}-\frac{1}{2\alpha}\frac{p_i^2}{E_p}u^{(0)}(p)e^{-ipx}.
\end{equation}
Thus we get
 \begin{equation}\label{c3}
  \hat{u}(p)=\bigg(1+a\alpha-\frac{a}{2}\frac{p_i^2}{E_p}\bigg)u^{(0)}(p).
\end{equation}
Now we consider Eq.(\ref{d7}) and by separating it in to $a$ independent and $a$ dependent coefficient terms, we get
\begin{equation}\label{c4}
 \Big(i\gamma^{\mu}\partial_{\mu}+m\Big)v^{(0)}(p)e^{ipx}=0,~~
 \Big(i\alpha\gamma^{\mu}\partial_{\mu}+\alpha m\Big)v^{(1)}(p)e^{ipx}+\frac{1}{2}\gamma^0\partial_i^2v^{(0)}e^{ipx}=0.
\end{equation}
We follow the similar steps as we did to find $\hat{u}(p)$ and obtain $v^{(1)}(p)$ and $\hat{v}(p)$ as
\begin{equation}\label{c4a}
 v^{(1)}(p)=\bigg(1-\frac{1}{2\alpha}\frac{p_i^2}{E_p}\bigg)v^{(0)}(p),
\end{equation}
and
\begin{equation}\label{c5}
 \hat{v}(p)=\bigg(1+a\alpha-\frac{a}{2}\frac{p_i^2}{E_p}\bigg)v^{(0)}(p).
\end{equation}
We consider the $\kappa$-deformed version of the unequal time anti-commutation relation between $\kappa$-deformed Dirac field operator and its adjoint as given in Eq.(\ref{a5})
\begin{equation}\label{c6}
 [\hat{\psi}(x),\bar{\hat{\psi}}(x')]_+=i\hat{d}(\partial)\hat{\Delta}(x-x')
\end{equation}
For a fermionic theory we assume the anti-commutation relations between deformed creation and anihilation operators in the usual form, i.e.,.,
\begin{equation}\label{c8}
\begin{split}
 [\hat{a}_s(p),\hat{a}_{s'}(p')]_+&=[\hat{b}_s(p),\hat{b}_{s'}(p')]_+=[\hat{a}^{\dagger}_{s}(p),\hat{a}^{\dagger}_{s'}(p')]_+=[\hat{b}^{\dagger}_s(p),\hat{b}^{\dagger}_{s'}(p')]_+=0, \\
[\hat{a}_s(p),\hat{a}^{\dagger}_{s'}(p')]_+&=[\hat{b}_s(p),\hat{b}^{\dagger}_{s'}(p')]_+=\delta_{ss'}\delta ^3(p-p').
\end{split}
\end{equation}
We now evaluate the LHS of Eq.(\ref{c6}) after substituting Eq.(\ref{c3}) and Eq.(\ref{c5}) in Eq.(\ref{decom1}) and Eq.(\ref{decom2}), i.e., 
\begin{equation}\label{c7}
\begin{split}
 [\hat{\psi}(x),\bar{\hat{\psi}}(x')]_+=&\int \frac{d^3p~d^3p'}{(2\pi)^3\sqrt{2E_p2E_{p'}}}\sum_{s,s'=1,2}\bigg\{[\hat{a}_s(p),\hat{a}^{\dagger}_{s'}(p')]_+\Big(\tilde{u}_s^{(0)}(p,x)\bar{\tilde{u}}_{s'}^{(0)}(p',x')\\+&a\alpha\big(\tilde{u}_s^{(0)}(p,x)\bar{\tilde{u}}_{s'}^{(1)}(p',x')+\tilde{u}_s^{(1)}(p,x)\bar{\tilde{u}}_{s'}^{(0)}(p',x')\big)\Big)\\
+&[\hat{b}_s(p),\hat{b}^{\dagger}_{s'}(p')]_+\Big(\tilde{v}_s^{(0)}(p,x)\bar{\tilde{v}}_{s'}^{(0)}(p',x')+a\alpha\big(\tilde{v}_s^{(0)}(p,x)\bar{\tilde{v}}_{s'}^{(1)}(p',x')+\tilde{v}_s^{(1)}(p,x)\bar{\tilde{v}}_{s'}^{(0)}(p',x')\big)\Big)\bigg\}.
\end{split}
\end{equation}
(here we have defined $\tilde{u}(p,x)=u(p)e^{-ipx}$ and $\tilde{v}(p,x)=v(p)e^{ipx}$). 
Using Eq.(\ref{c8}) in Eq.(\ref{c7}), we get
\begin{equation}\label{c9}
\begin{split}
 [\hat{\psi}(x),\bar{\hat{\psi}}(x')]_+=\int \frac{d^3p}{(2\pi)^32E_p}\sum_{s=1,2}\bigg\{{u}_s^{(0)}(p)\bar{{u}}_{s}^{(0)}(p)e^{-ip(x-x')}+a\alpha\big({u}_s^{(0)}(p)\bar{{u}}_{s}^{(1)}(p)+{u}_s^{(1)}(p)\bar{{u}}_{s}^{(0)}(p)\big)e^{-ip(x-x')}\\+{v}_s^{(0)}(p)\bar{{v}}_{s}^{(0)}(p)e^{ip(x-x')}+a\alpha\big({v}_s^{(0)}(p)\bar{{v}}_{s}^{(1)}(p)+{v}_s^{(1)}(p)\bar{{v}}_{s}^{(0)}(p)\big)e^{ip(x-x')}\bigg\}.
\end{split}
\end{equation}
Using Eq.(\ref{c2a}) and Eq.(\ref{c4a}), we evaluate ${u}_s^{(0)}(p)\bar{{u}}_{s}^{(1)}(p),~{u}_s^{(1)}(p)\bar{{u}}_{s}^{(0)}(p),~{v}_s^{(0)}(p)\bar{{v}}_{s}^{(1)}(p),~{v}_s^{(1)}(p)\bar{{v}}_{s}^{(0)}(p)$ as
\begin{equation}\label{z1}
 {u}_s^{(0)}(p)\bar{{u}}_{s}^{(1)}(p)={u}_s^{(0)}(p)\bar{{u}}_{s}^{(0)}(p)\bigg(1-\frac{1}{2\alpha}\frac{p_i^2}{E_p}\bigg),
\end{equation}
\begin{equation}\label{z2}
 {u}_s^{(1)}(p)\bar{{u}}_{s}^{(0)}(p)={u}_s^{(0)}(p)\bar{{u}}_{s}^{(0)}(p)\bigg(1-\frac{1}{2\alpha}\frac{p_i^2}{E_p}\bigg),
\end{equation}
\begin{equation}\label{z3}
 {v}_s^{(0)}(p)\bar{{v}}_{s}^{(1)}(p)={v}_s^{(0)}(p)\bar{{v}}_{s}^{(0)}(p)\bigg(1-\frac{1}{2\alpha}\frac{p_i^2}{E_p}\bigg),
\end{equation}
\begin{equation}\label{z4}
 {v}_s^{(1)}(p)\bar{{v}}_{s}^{(0)}(p)={v}_s^{(0)}(p)\bar{{v}}_{s}^{(0)}(p)\bigg(1-\frac{1}{2\alpha}\frac{p_i^2}{E_p}\bigg).
\end{equation}
Substituting Eq.(\ref{z1}), Eq.(\ref{z2}), Eq.(\ref{z3}) and Eq.(\ref{z4}) in Eq.(\ref{c9}) and using the completeness relation i.e., $\sum_{s=1,2}u^{(0)}_s\bar{u}^{(0)}_s=\slashed{p}-m$ and $\sum_{s=1,2}v^{(0)}_s\bar{v}^{(0)}_s=\slashed{p}+m$, we get
\begin{equation}\label{c10}
\begin{split}
 [\hat{\psi}(x),\bar{\hat{\psi}}(x')]_+=&i\big(1+2a\alpha\big)(i\slashed{\partial}-m)\Delta(x-x')-2a\big(i\slashed{\partial}-m\big)\int\frac{d^3p}{(2\pi)^3}\frac{p_i^2}{(2E_p)^2}\Big(e^{-ip(x-x')}-e^{ip(x-x')}\Big).
\end{split}
\end{equation}
We notice that first terms of the Eq.(\ref{z1}), Eq.(\ref{z2}), Eq.(\ref{z3}) and Eq.(\ref{z4}) add up and contribute the $i2a\alpha\big(i\slashed{\partial}-m\big)\Delta(x-x')$ term present in the RHS of Eq.(\ref{c10}). Similarly the second term of the Eq.(\ref{z1}), Eq.(\ref{z2}), Eq.(\ref{z3}) and Eq.(\ref{z4}) add up and contribute the $\alpha$ independent correction term in Eq.(\ref{c10}) and this last integral vanishes as the integrand being an odd function. Now we compute the RHS of Eq.(\ref{c6}) as
\begin{equation}\label{c11}
\begin{split}
 i\hat{d}(\partial)\hat{\Delta}(x-x')=&i\Big(i\slashed{\partial}+\frac{a}{2}\gamma^0\partial_i^2-m\Big)\Big(\Delta(x-x')+a\Delta ^{(1)}(x-x')\Big)
\\
=&i(i\slashed{\partial}-m)\Delta(x-x')+ia(i\slashed{\partial}-m)\Delta^{(1)}(x-x')+i\frac{a}{2}\gamma^0\int\frac{d^3p}{(2\pi)^32E_p}p_i^2\Big(e^{-ip(x-x')}-e^{ip(x-x')}\Big).
\end{split}
\end{equation}
In the above equation, $\Delta^{(1)}(x-x')$ is the first order correction to $\hat{\Delta}(x-x')$, which is to be calculated. Here too the last integral vanishes as the integrand being an odd function. Now we compare this with Eq.(\ref{c10}) and find that $\Delta^{(1)}(x-x')=2\alpha\Delta(x-x')$ for the undeformed oscillator algebra given in Eq.(\ref{c8}). Hence we find that $\hat{\Delta}(x-x')$ is same as that obtained for the $\kappa$-deformed Klein-Gordon equation \cite{vishnu1}. Thus the anti-commutation relation between deformed field operator and its adjoint becomes
\begin{equation}\label{c12}
 [\hat{\psi}(x),\bar{\hat{\psi}}(x')]_+=i\big(1+2a\alpha\big)(i\slashed{\partial}-m)\Delta(x-x').
\end{equation}
Now we consider a situtation where the anti-commutation relation between creation and annihilation operators is allowed to be deformed, and this deformation is vaild up to first order in $a$. i.e., we assume the deformed oscillator algebra to be of the form
\begin{equation}\label{c13}
\begin{split}
 [\hat{a}_s(p),\hat{a}_{s'}(p')]_+&=[\hat{b}_s(p),\hat{b}_{s'}(p')]_+=[\hat{a}^{\dagger}_{s}(p),\hat{a}^{\dagger}_{s'}(p')]_+=[\hat{b}^{\dagger}_s(p),\hat{b}^{\dagger}_{s'}(p')]_+=0, \\
[\hat{a}_s(p),\hat{a}^{\dagger}_{s'}(p')]_+&=[\hat{b}_s(p),\hat{b}^{\dagger}_{s'}(p')]_+=h(a)\delta_{ss'}\delta ^3(p-p').
\end{split}
\end{equation}
Note that in the above, $h(a)$ is an unknown function in $a$. Using this deformed oscillator algebra in Eq.(\ref{c7}) and following the above steps, we get the, unequal time anti-commutation relation between deformed field operator and its adjoint in the form
\begin{equation}\label{c14}
 [\hat{\psi}(x),\bar{\hat{\psi}}(x')]_+=ih(a)\big(1+2a\alpha\big)(i\slashed{\partial}-m)\Delta(x-x').
\end{equation}
We now choose $h(a)$ such that the anti-commutation relation between deformed field operator and its adjoint becomes undeformed, up to first order in $a$. This sets $h(a)=1-2a\alpha$. Thus the undeformed anti-commutation relation between deformed field operator and its adjoint takes the form
\begin{equation}\label{c15}
 [\hat{\psi}(x),\bar{\hat{\psi}}(x')]_+=i(i\slashed{\partial}-m)\Delta(x-x').
\end{equation}
The $\kappa$-deformed anti-commutation relations between deformed creation and anihilation operator are thus given by
\begin{equation}\label{c16}
\begin{split}
 [\hat{a}_s(p),\hat{a}_{s'}(p')]_+&=[\hat{b}_s(p),\hat{b}_{s'}(p')]_+=[\hat{a}^{\dagger}_{s}(p),\hat{a}^{\dagger}_{s'}(p')]_+=[\hat{b}^{\dagger}_s(p),\hat{b}^{\dagger}_{s'}(p')]_+=0, \\
[\hat{a}_s(p),\hat{a}^{\dagger}_{s'}(p')]_+&=[\hat{b}_s(p),\hat{b}^{\dagger}_{s'}(p')]_+=(1-2a\alpha)\delta_{ss'}\delta ^3(p-p').
\end{split}
\end{equation}
We observe that the deformation factor appearing in the deformed oscillator corresponding to $\kappa$-deformed Dirac field is same as that of the deformation factor appearing in the deformed oscillator algebra of $\kappa$-deformed Klein-Gordon field \cite{vishnu1}.

\section{$\kappa$-deformed conserved currents}

 In this section we first construct the $\kappa$-deformed version of the Gamma operator, given in Eq.(\ref{a6}), using the Gamma operator and then we write down the general expression for the $\kappa$-deformed conserved current using this deformed Gamma operator, $\hat{\Gamma}_{\mu}(\partial,-\overleftarrow{\partial})$. Using this expression for the deformed conserved currents we then derive the conserved currents corresponding to global phase transformation as well as those corresponding to discrete symmetry transformations of $\kappa$-Dirac equation, valid up to first order in $a$. 

The $\hat{\Gamma}_{\mu}(\partial,-\overleftarrow{\partial})$ operator for the $\kappa$-deformed Dirac field is constructed from $\hat{\Lambda}(\partial)$, by substituting $\hat{\Lambda}(\partial)$ in Eq.(\ref{a6}) and its explicit form, valid up to first order in $a$ is given by
\begin{equation}\label{e1}
 \hat{\Gamma}^{\mu}(\partial,-\overleftarrow{\partial})=i\gamma^{\mu}+\frac{a}{2}\gamma^0\delta^{\mu i}(\partial_i-\overleftarrow{\partial}_i).
\end{equation}
It is to be noted that the second term in the RHS is solely due to the contribution from the $\kappa$-deformation, valid up to first order in $a$. To show that the above operator satisfy the deformed version of Eq.(\ref{iden}), we first notice that the operator defined above satisfies
\begin{equation}\label{g1a}
 \left(\partial_{\mu}+\overleftarrow{\partial}_{\mu}\right)\hat{\Gamma}_{\mu}(\partial,-\overleftarrow{\partial})=i\gamma^{\mu}\left(\partial_{\mu}+\overleftarrow{\partial}_{\mu}\right)+\frac{a}{2}\gamma^0\delta^{\mu i}\left(\partial_{\mu}+\overleftarrow{\partial}_{\mu}\right)\left(\partial_{i}-\overleftarrow{\partial}_{i}\right).
\end{equation}
Next from Eq.(\ref{lambda}), we obtain the expression
\begin{equation}\label{g2a}
 \hat{\Lambda}(\partial)-\hat{\Lambda}(-\overleftarrow{\partial})=i\gamma^{\mu}\left(\partial_{\mu}+\overleftarrow{\partial}_{\mu}\right)+\frac{a}{2}\gamma^0\delta^{\mu i}\left(\partial_{\mu}+\overleftarrow{\partial}_{\mu}\right)\left(\partial_{i}-\overleftarrow{\partial}_{i}\right).
\end{equation}
Comparing Eq.(\ref{g1a}) and Eq.(\ref{g2a}) we find that $\left(\partial_{\mu}+\overleftarrow{\partial}_{\mu}\right)\hat{\Gamma}_{\mu}(\partial,-\overleftarrow{\partial})=\hat{\Lambda}(\partial)-\hat{\Lambda}(-\overleftarrow{\partial})$,
 and deformed operator $\hat {\Gamma}^\mu$ defined above satisfies Eq.(\ref{iden})  and hence this deformed Gamma operator is consistent and unique for constructing the deformed conserved currents.
 
Now using Eq.(\ref{conser}), we write the expression for the deformed conserved current as
\begin{equation}\label{e15}
 \hat{J}^{\mu}=\bar{\hat{\psi}}(x)\hat{\Gamma}^{\mu}(\partial,-\overleftarrow{\partial})\delta\hat{\psi}(x)+h.c,
\end{equation}
where we have taken $G=\bar{\hat{\psi}}(x)$ and $F=\delta\hat{\psi}(x)$, the variation of the field under the transformation of interest.

Under an infintesimal global phase transformation, we have $\hat{\psi}(x)\to\hat{\psi}'(x)$, where $\hat{\psi}'(x)=e^{-i\theta}\hat{\psi}(x)$, and hence $\delta\hat{\psi}(x)=-i\theta\hat{\psi}(x)$. Thus the corresponding deformed 
conserved current, using Eq.(\ref{e15}) is obtained as 
\begin{equation}\label{e*3}
 \hat{J}^{\mu}=-i\bar{\hat{\psi}}(x)\Big( i\gamma^{\mu}+\frac{a}{2}\gamma^0\delta^{\mu i}(\partial_i-\overleftarrow{\partial}_i) \Big)\hat{\psi}(x)+h.c.
\end{equation}
Next we construct the $\kappa$-deformed number operator as $\hat{N}=\int d^3x \hat{J}^0(x)$, i.e.,
\begin{equation}\label{e*}
\begin{split}
 \hat{N}=&\int d^3x \bar{\hat{\psi}}(x)\gamma^0\hat{\psi}(x)\\
=&\int d^3x \frac{d^3p}{\sqrt{(2\pi)^3 2E_p}}\frac{d^3p'}{\sqrt{(2\pi)^3 2E_{p'}}}\Big(\hat{a}^{\dagger}(p)\bar{\hat{u}}(p)e^{ipx}+\hat{b}(p)\bar{\hat{v}}(p)e^{-ipx}\Big)\gamma^0\Big(\hat{a}(p')\hat{u}(p')e^{-ip'x}+\hat{b}^{\dagger}(p'){\hat{v}}(p')e^{ip'x}\Big)\\
=&\int d^3x \frac{d^3p}{\sqrt{(2\pi)^3 2E_p}}\frac{d^3p'}{\sqrt{(2\pi)^3 2E_{p'}}}\Big(\hat{a}^{\dagger}(p)\hat{a}(p')\hat{u}^{\dagger}(p)\hat{u}(p')e^{i(p-p')x}+\hat{a}^{\dagger}(p)\hat{b}^{\dagger}(p')\hat{u}^{\dagger}(p)\hat{v}(p')e^{i(p+p')x}\\&+\hat{b}(p)\hat{a}(p')\hat{v}^{\dagger}(p)\hat{u}(p')e^{-i(p+p')x}+\hat{b}(p)\hat{b}^{\dagger}(p')\hat{v}^{\dagger}(p)\hat{v}(p')e^{-i(p-p')x}\Big)
\end{split}
\end{equation} 
Now substituting Eq.(\ref{c3}) and Eq.(\ref{c5}) in Eq.(\ref{e*}) and using the relations ${u}^{(0)\dagger}_s(p){v}^{(0)}_{s'}(-p)={v}^{(0)\dagger}_s(p){u}^{(0)}_{s'}(-p)=0$ and ${u}^{(0)\dagger}_s(p){u}^{(0)}_{s'}(p')={v}^{(0)\dagger}_s(p){v}^{(0)}_{s'}(p')=\delta_{ss'}2E_p$ and symplifying, we get
\begin{equation}\label{e*1}
 \hat{N}=\int \frac{d^3p}{(2\pi)^3}\Big(1+2a\alpha-a\frac{p^2}{E_p}\Big)\Big(\hat{a}^{\dagger}(p)\hat{a}(p)+\hat{b}(p)\hat{b}^{\dagger}(p)\Big)
\end{equation}
We denote $\hat{N}_a(p)=\hat{a}^{\dagger}(p)\hat{a}(p)$ and $\hat{N}_b(p)=\hat{b}^{\dagger}(p)\hat{b}(p)$ and after normal ordering we get the deformed number operator, valid up to first order in $a$, as
\begin{equation}\label{e*2}
 :\hat{N}:=\int \frac{d^3p}{(2\pi)^3}\Big(1+2a\alpha-a\frac{p^2}{E_p}\Big)\Big(\hat{N}_a(p)-\hat{N}_b(p)\Big).
\end{equation}
Here we see that the deformed number operator gets modified by a term that depends only on $a\alpha$ as well as a mass-dependent term, $-a\frac{p^2}{E_p}$. In the commutative limit, i.e.,, $a\to 0$, we recover the Dirac number operator as, $:N:=\int d^3p \Big({N}_a(p)-{N}_b(p)\Big)$. Note that the non-commutative corrections to number operator may have implications in Unruh effect.

We now consider the discrete symmetry transformation of this equation and construct corresponding conserved currents. For this, we start with the $\kappa$-deformed Dirac equation, valid upto first order in $a$, given in Eq.(\ref{d3}), re-written as
\begin{equation}\label{e3}
 \Big(i\gamma^0\partial_0+i\gamma^i\partial_i+\frac{a}{2}\gamma^0\partial_i^2+m\Big)\hat{\psi}(x_i,t)=0.
\end{equation}   
We now consider the invariance of this equation under parity, time-reversal and charge conjugation, respectively \cite{lur}.

\subsection{Parity}

Under the parity transformation, we have $x_i\to -x_i,~t\to t,~\partial_i\to-\partial_i$ and $\partial_0\to\partial_0$, hence Eq.(\ref{e3}) becomes
\begin{equation}\label{e4}
 \Big(i\gamma^0\partial_0-i\gamma^i\partial_i+\frac{a}{2}\gamma^0\partial_i^2+m\Big)\hat{\psi}(-x_i,t)=0
\end{equation} 
Let there be a matrix $\mathcal{P}$ such that
\begin{equation}\label{e5}
 \mathcal{P}\Big(i\gamma^0\partial_0-i\gamma^i\partial_i+\frac{a}{2}\gamma^0\partial_i^2+m\Big)\mathcal{P}^{-1}\mathcal{P}\hat{\psi}(-x_i,t)=0
\end{equation} 
where $\mathcal{P}\Big(i\gamma^0\partial_0-i\gamma^i\partial_i+\frac{a}{2}\gamma^0\partial_i^2+m\Big)\mathcal{P}^{-1}$ is $\hat{\Lambda}(\partial)$, given in Eq.(\ref{lambda}). This is possible if and only if  $\mathcal{P}\hat{\psi}(-x_i,t)=\hat{\psi}_p(x_i,t)$  satisfy the $\kappa$-deformed Dirac equation. We find that $\mathcal{P}=\gamma^0$, satisfy above requirement and it is the same as that in the commutative case. Thus parity is a symmetry for $\kappa$-deformed Dirac equation. 

Now we determine the deformed conserved current corresponding to parity symmetry for the $\kappa$-deformed Dirac equation, valid up to first order in $a$.  Here the $\hat{\Lambda}(\partial)$ operator is the same as that in Eq.
(\ref{lambda}), so we use the corresponding $\hat{\Gamma}_{\mu}(\partial,-\overleftarrow{\partial})$ defined in Eq.(\ref{j1}) for constructing deformed current for parity symmetry. Thus the deformed conserved current associated with parity, using Eq.(\ref{e15}), by taking $\delta\hat{\psi}(x)=\hat{\psi}_p(x_i,t)$ (see appendix for details) is
\begin{equation}\label{e6}
\begin{split}
 \hat{J}^{\mu}=&\bar{\hat{\psi}}(x_i,t)\hat{\Gamma}^{\mu}(\partial,-\overleftarrow{\partial})\mathcal{P}\hat{\psi}(-x_i,t)+h.c\\
=&i\bar{\hat{\psi}}(x_i,t)\gamma^{\mu}\hat{\psi}_p(x_i,t)+\frac{a}{2}\gamma^0\delta^{\mu i}\bar{\hat{\psi}}(x_i,t)(\partial_i-\overleftarrow{\partial}_i)\hat{\psi}_p(x_i,t)+h.c.
\end{split}
\end{equation}
From Eq.(\ref{e6}) we notice that the second term on RHS is purely due to the non-commutative correction and this term is present only in the spatial part of the deformed conserved current. Now we obtain the explicit form of the conserved charge corresponding to parity symmetry, by substituting Eq.(\ref{c3}) and Eq.(\ref{c5}) in above as
\begin{equation}\label{pcharge}
\begin{split}
 \hat{Q}_p=&\int\frac{dp^3}{(2\pi)^3}\Big(1+2a\alpha-a\frac{p^2}{E_p}\Big)\Big(\hat{a}^{\dagger}(p)\hat{a}(-p)+\hat{b}^{\dagger}(p)\hat{b}(-p)\Big)
\end{split}
\end{equation}
Here we see that conserved charge corresponding to parity symmetry in $\kappa$-deformed space-time picks up two correction terms, of which first one is mass independent and the second one is mass-dependent. This non-commutative correction factor is exaclty the same as non-commutative correction term present in the deformed number operator Eq.(\ref{e*2}).  The dependence of the conserved charge on mass of the particle will have implications
in neutrino physics. This is a novel feature of physics in the $\kappa$-deformed space-time.

Note that the conserved charge in $\kappa$-deformed space-time has the same form as that in commutative space-time and the $a$-dependent corrections come only from the deformed field operator and its adjoint. This is due to the fact that the parity operator is the same in both commutative and $\kappa$-deformed space-time. In the commutative limit, the conserved current corresponding to the parity symmetry of Dirac equation reduces to $J^{\mu}=i\bar{{\psi}}(x_i,t)\gamma^{\mu}{\psi}_p(x_i,t)+h.c$ \cite{lur}.

\subsection{Time-reversal} 

Under the time-reversal we have $x_i\to x_i,~t\to -t,~\partial_i\to\partial_i$ and $\partial_0\to-\partial_0$, hence Eq.(\ref{e3}) becomes
\begin{equation}\label{e7}
 \Big(-i\gamma^0\partial_0+i\gamma^i\partial_i+\frac{a}{2}\gamma^0\partial_i^2+m\Big)\hat{\psi}(x_i,-t)=0
\end{equation} 
and taking its comlpex conjugate we get
\begin{equation}\label{e8}
 \Big(i\gamma^{0*}\partial_0-i\gamma^{i*}\partial_i+\frac{a}{2}\gamma^{0*}\partial_i^2+m\Big)\hat{\psi}^*(x_i,-t)=0.
\end{equation} 
Let us consider a matrix $\mathcal{T}$ such that is obeys,
\begin{equation}\label{e9}
 \mathcal{T}\Big(i\gamma^{0*}\partial_0-i\gamma^{i*}\partial_i+\frac{a}{2}\gamma^{0*}\partial_i^2+m\Big)\mathcal{T}^{-1}\mathcal{T}\hat{\psi}^*(x_i,-t)=0.
\end{equation}
where $\mathcal{T}\Big(i\gamma^{0*}\partial_0-i\gamma^{i*}\partial_i+\frac{a}{2}\gamma^{0*}\partial_i^2+m\Big)\mathcal{T}^{-1}$ is $\hat{\Lambda}(\partial)$, given in Eq.(\ref{e4}). This constraint is satisfied if $\mathcal{T}\hat{\psi}^*(x_i,-t)=\hat{\psi}_T(x_i,t)$  satisfy the $\kappa$-deformed Dirac equation. By comparing Eq.(\ref{e9}) with $\hat{\Lambda}(\partial)$ we find $\mathcal{T}\gamma^{0*}\mathcal{T}^{-1}=\gamma^0$ and $\mathcal{T}\gamma^{i*}\mathcal{T}^{-1}=-\gamma^i$ and hence we obtain $\mathcal{T}=i\gamma^1\gamma^3$. Thus time-reversal operator is the same as that in commutative space-time. We find that time-reversal is also a symmetry for $\kappa$-deformed Dirac equation.

Now we obtain the deformed conserved current corresponding to the time-reversal symmetry of the $\kappa$-deformed Dirac equation, valid up to first order in $a$. We see that the $\hat{\Lambda}(\partial)$ operator used here is same as that in Eq.(\ref{lambda}), so we use the corresponding $\hat{\Gamma}_{\mu}(\partial,-\overleftarrow{\partial})$ defined in Eq.(\ref{j1}) for constructing the deformed current for time-reversal symmetry. Hence the deformed conserved current for time-reversal is defined as \cite{lur}.
\begin{equation}\label{e10}
\begin{split}
 \hat{J}^{\mu}=&\bar{\hat{\psi}}(x_i,t)\hat{\Gamma}^{\mu}(\partial,-\overleftarrow{\partial})\mathcal{T}\hat{\psi}^*(x_i,-t)+h.c\\
=&i\bar{\hat{\psi}}(x_i,t)\gamma^{\mu}\hat{\psi}_T(x_i,t)+\frac{a}{2}\gamma^0\delta^{\mu i}\bar{\hat{\psi}}(x_i,t)(\partial_i-\overleftarrow{\partial}_i)\hat{\psi}_T(x_i,t)+h.c.
\end{split}
\end{equation}
From Eq.(\ref{e10}) we observe that the second term on RHS is purely due to the contribution from the  $\kappa$-deformed correction and this term is present only in the spatial part of the $\kappa$-deformed conserved current. Now we obtain the explicit form of the conserved charge corresponding to time-reversal symmetry, by substituting Eq.(\ref{c3}) and Eq.(\ref{c5}) in above, as
\begin{equation}\label{tcharge}
\begin{split}
\hat{Q}_t=&\int\frac{d^3p}{(2\pi)^3}\Big(1+2a\alpha-a\frac{p^2}{E_p}\Big)\Big(\hat{a}^{\dagger}(p)\hat{a}^*(-p)+\hat{b}(p)\hat{b}^T(-p)\Big).
\end{split}
\end{equation}
We observe that the conserved charge corresponding to time-reversal symmetry in $\kappa$-deformed space-time also posses same $a$ dependent correction terms as in the case of parity and global phase transformation symmetries. Here too the second term depend on the mass of the particle.  This will also have implications in particle physics as the conserved charge depends on the mass, which is a novel feature of the $\kappa$-deformed Dirac 
equation.

Note that the $a$-dependent corrections come only from the deformed field operator and its adjoint. This is consistent with the fact that the time-reversal operator in $\kappa$ space-time as in the commutative case. In the commutative limit, the conserved current corresponding to the time-reversal symmetry of Dirac equation reduces to $J^{\mu}=i\bar{{\psi}}(x_i,t)\gamma^{\mu}{\psi}_T(x_i,t)+h.c$ \cite{lur}.

\subsection{Charge conjugation}
To study the symmetry property of $\kappa$-deformed Dirac equation under charge conjugation, we introduce a minimal coupling of electron with electromagnetic field $A_{\mu}$ , by replacing $i\partial_{\mu}$ in the $\kappa$-deformed Dirac equation with $i\partial_{\mu}\to i\partial_{\mu}+eA_{\mu}$. Hence the $\hat{\Lambda}(\partial)$ operator becomes
\begin{equation}\label{j1}
 \hat{\Lambda}_{c}(\partial)=\gamma^{\mu}\Big(i\partial_{\mu}+eA_{\mu}\Big)-\frac{a}{2}\gamma^0\Big(i\partial_{i}+eA_{i}\Big)^2+m
\end{equation}
satisfying
\begin{equation}\label{j2}
 \hat{\Lambda}_{c}(\partial)\hat{\psi}(x_i,t)=0.
\end{equation}
Under the charge conjugation $e\to-e$, and we get the deformed equation of motion for the anti-particle as
\begin{equation}\label{j3}
 \bigg[\gamma^{\mu}\Big(i\partial_{\mu}-eA_{\mu}\Big)-\frac{a}{2}\gamma^0\Big(i\partial_{i}-eA_{i}\Big)^2+m\bigg]\hat{\psi}_c(x_i,t)=0.
\end{equation}
Substituting Eq.(\ref{j1}) in Eq.(\ref{j2}) and taking the complex conjugate, we obtain the following equation
\begin{equation}\label{j4}
 \bigg[\gamma^{*\mu}\Big(-i\partial_{\mu}+eA_{\mu}\Big)-\frac{a}{2}\gamma^{*0}\Big(-i\partial_{i}+eA_{i}\Big)^2+m\bigg]\hat{\psi}^*(x_i,t)=0.
\end{equation}
Let us consider a $\mathcal{C}$ martix such that
\begin{equation}\label{j5}
 \mathcal{C}\bigg[-\gamma^{*\mu}\Big(i\partial_{\mu}-eA_{\mu}\Big)-\frac{a}{2}\gamma^{*0}\Big(i\partial_{i}-eA_{i}\Big)^2+m\bigg]\mathcal{C}^{-1}\mathcal{C}\hat{\psi}^*(x_i,t)=0,
\end{equation}
and one demands  Eq.(\ref{j5}) to be equivalent to Eq.(\ref{j3}). Thus we find that $\hat{\psi}_c(x_i,t)=\mathcal{C}\hat{\psi}^*(x_i,t)$  and $\mathcal{C}$ has to satisfy other conditions(see below) as well.

Now we construct the $\hat{\Gamma}^{\mu}_c(\partial,-\overleftarrow{\partial})$ operator corresponding to the $\hat{\Lambda}(\partial)$, defined in eq.(\ref{j1}), using Eq.(\ref{a6}), valid up to first order in $a$ as
\begin{equation}\label{j21}
 \hat{\Gamma}^{\mu}_c(\partial,-\overleftarrow{\partial})=i\left(\gamma^{\mu}-a\gamma^0eA_i\delta^{\mu i}\right)+\frac{a}{2}\gamma^0\delta^{\mu i}\left(\partial_i-\overleftarrow{\partial}_i\right).
\end{equation}
We use this $\hat{\Gamma}^{\mu}_c$ operator and construct the conserved current corresponding to the charge conjugation symmetry for the $\kappa$-deformed Dirac equation, valid up to first order in $a$, as
\begin{equation}\label{j6}
 \hat{J}^{\mu}_c=\bar{\hat{\psi}}(x_i,t)\hat{\Gamma}^{\mu}_c(\partial,-\overleftarrow{\partial})\delta\hat{\psi}(x_i,t)+h.c,
\end{equation}
where we have $\delta\hat{\psi}(x_i,t)=\hat{\psi}_c(x_i,t)=\mathcal{C}\hat{\psi}^*(x_i,t)$ and $\hat{\Gamma}^\mu$ has to satisfy Eq.(\ref{iden}). To check this, using the $\hat{\Gamma}^{\mu}_c(\partial,-\overleftarrow{\partial})$ equation, i.e., Eq.(\ref{j21}) we fine
\begin{equation}\label{j22}
 \left(\partial_{\mu}+\overleftarrow{\partial}_{\mu}\right)\hat{\Gamma}^{\mu}_c(\partial,-\overleftarrow{\partial})=i\gamma^{\mu}\left(\partial_{\mu}+\overleftarrow{\partial}_{\mu}\right)-ia\gamma^0eA_i\left(\partial_i+\overleftarrow{\partial}_i\right)+\frac{a}{2}\gamma^0\left(\partial^2_i-\overleftarrow{\partial}^2_i\right).
\end{equation}  
Then from Eq.(\ref{j1}) we obtain
\begin{equation}\label{j23}
 \hat{\Lambda}_c(\partial)-\hat{\Lambda}_c(-\overleftarrow{\partial})=i\gamma^{\mu}\left(\partial_{\mu}+\overleftarrow{\partial}_{\mu}\right)-ia\gamma^0eA_i\left(\partial_i+\overleftarrow{\partial}_i\right)+\frac{a}{2}\gamma^0\left(\partial^2_i-\overleftarrow{\partial}^2_i\right)+i\gamma^0e\left(\overleftarrow{\partial}_iA_i\right)-i\gamma^0e\left(\partial_iA_i\right).
\end{equation}
Comapring Eq.(\ref{j22}) and Eq.(\ref{j23}) we find that, $\left(\partial_{\mu}+\overleftarrow{\partial}_{\mu}\right)\hat{\Gamma}^{\mu}_c(\partial,-\overleftarrow{\partial})\neq \hat{\Lambda}_c(\partial)-\hat{\Lambda}_c(-\overleftarrow{\partial})$, the identity given in Eq.(\ref{iden}) is violated by the corresponding $\hat{\Gamma}^{\mu}_c$ operator. Thus one can not construct a consistent $\hat{\Gamma}^\mu$ operator and hence 
the corresponding conserved current in Eq,(\ref{j6}) does not exist. This re-confirm the violation of charge conjugation symmetry by particles satisfying $\kappa$-deformed Dirac equation, using a different approach. 

In \cite{sivakumar} it has been shown that the charge conjugation is not a symmetry of the $\kappa$-deformed Dirac equation by showing the charge conjugation matrix, $\mathcal{C}$ does not exist, using the conditions $\mathcal{C}\gamma^{*\mu}\mathcal{C}^{-1}=-\gamma^{\mu},\mathcal{C}\gamma^{*0}\mathcal{C}^{-1}=\gamma^{0}, \mathcal{C}\mathcal{C}^{-1}=\mathds{I}$ obtained by comparing Eq.(\ref{j3}) and Eq.(\ref{j5}). Here we obtain the same result using an alternative approach.

 \section{Conclusions}
 
In this paper, starting with the $\kappa$-deformed Dirac field, satisfying $\kappa$-deformed Dirac equation\cite{sivakumar}, valid up to first order in $a$, for a particular choice of realisation, we have quantised the Dirac spinor in $\kappa$-deformed space-time, following the method developed in \cite{yt,ume,ytaka}. The $\kappa$-deformed Dirac equation has been constructed \cite{sivakumar} from the Dirac derivative, which transforms as $4$-vector under undeformed $\kappa$-Poincare algebra. It is to be noted that we quantised the $\kappa$-deformed Dirac spinor using the equations of motion alone, without using its Lagrangian.

Using the $\kappa$-deformed Dirac equation and its adjoint, we have constructed the $\hat{\Lambda}(\partial)$ operator and Klein-Gordon divisor, $\hat{d}(\partial)$, for the deformed Dirac field in the $\kappa$-space-time, valid up to first order in $a$. We further used the $\hat{\Lambda}(\partial)$ operator and derived the $\hat{\Gamma}^{\mu}(\partial,-\overleftarrow{\partial})$ operator, valid up to first order in $a$, in the $\kappa$-deformed space-time. The $\kappa$-deformed Dirac equation is then solved perturbatively obtaining solutions, valid up to first order in $a$. Using this solution, the unequal-time anti-commutation relation between the deformed Dirac field and its adjoint field, valid up to first order in $a$ is derived. The expression for the anti-commutation relation is shown to be modified by the space-time deformation and in the limit $a\to 0$, we get back the well-known result. This modification in anti-commutator is given in terms of the Green's function, $\hat{\Delta}(x-x')$ and we found that $\hat{\Delta}(x-x')$ is exactly the same as that obtained for $\kappa$-deformed Klein-Gordon equation \cite{vishnu1}. The undeformed oscillator algebra between the creation and annihilation operators gave rise to a deformed unequal-time anti-commutation relation between deformed Dirac field and its adjoint field, where $(1+2a\alpha)$ is the deformation factor. In \cite{vishnu1} we showed that undeformed oscillator algebra leads to deformed unequal time commutation relation between deformed Klein-Gordan field and its conjugate, with the same deformation factor $(1+2a\alpha)$. By demanding the unequal time anti-commutation relation between deformed field and its adjoint to be undeformed, we obtained a deformed oscillator algebra, where the deformation factor, i.e., $(1-2a\alpha)$, valid up to first order in $a$. This $\kappa$  deformed oscillator algebra will also modify Unruh effect as well as Hawking radiation. Thus we have seen that the deformed oscillator algebra is a novel feature of the $\kappa$-deformed theory. This deformation factor present in the deformed oscillator algebra is also the same as that of the deformation factor present in the corresponding algebra of $\kappa$-deformed Klein-Gordan field \cite{vishnu1}.

The deformed oscillator algebra derived here is different from the one derived in  \cite{ravi2} by demanding the compatibility between twisted flip-operator and action of Hopf algebra.  We have shown that the deformed algebra associated with $\kappa$-deformed Kelin-Gordon field has implications for Unruh effect as well as Hawking radiation\cite{vishnu1}. The deformed fermionic oscillator algebra corresponding to the $\kappa$-deformed Dirac field will also be expected to have implications in particle physics, especially in the $S$-matrix calculations and the scattering cross-section.

We have derived the $\kappa$-deformed number operator from the deformed conserved charge corresponding to the global phase transformation symmetry as in the commutative case \cite{ytaka}, and the number operator picks up a mass-dependent correction term in the $\kappa$-deformed space-time. We have also studied the discrete symmetry in the $\kappa$-deformed Dirac equation and using the $\hat{\Gamma}^{\mu}(\partial,-\overleftarrow{\partial})$, the expression for the deformed conserved currents and charges, valid up to first order in $a$, corresponding to parity and time-reversal symmetry have been derived, without referring to the $\kappa$-deformed Lagrangian. In both these cases, the conserved charges have the same operator form as that in the commutative case, and the $a$-dependent correction is coming only from the $a$-dependent corrections of the deformed field and its adjoint. For global gauge transformation as well as for parity and time reversal, the $\kappa$-deformed corrections to the conserved quantities are of the same form. Of the two correction terms, one depends on the mass of the Dirac quanta, which is a novel result. This will have measurable consequences in particle physics. This is an important feature brought out by the approach of quantisation and construction of the conserved quantities employed here.  We also have shown that charge conjugation is not a symmetry of the $\kappa$-deformed Dirac equation by showing that $\hat{\Gamma}^{\mu}_c(\partial,-\overleftarrow{\partial})$ operator does not exist.

\section{Appendix A}

Here, we give a brief summary of the construction of conserved charges appropriate for parity and time-reversal symmetries of Dirac equation in commutative space-time \cite{lur}.

From the Dirac Lagrangian, $\mathcal{L}=\bar{\psi}(x)\big(i\slashed{\partial}+m\big)\psi(x)$, we obtain the conjugate momenta for $\psi(x)$ as $\pi_{\psi}=i\psi^{\dagger}$ satisfying the relation
\begin{equation}\label{g2}
 \{\psi(x_i,t),\pi_{\psi}(x'_i,t)\}_{P.B}=\delta^3(x_i-x'_i).
\end{equation}
The variation in $\psi(x)$, i.e., $\delta\psi(x)$ due to a symmetry transformation can be determined from the generator of the symmetry transformation using the relation given by
\begin{equation}\label{g3}
 \delta\psi(x)=\{\psi(x_i,t),Q\}_{P.B}
\end{equation}
The conserved current corresponding to the parity symmetry of the commutative Dirac equation is $J^{\mu}=i\bar{\psi}(x_i,t)\gamma^{\mu}\psi_p(x_i,t)$, where $\psi_p(x_i,t)=\gamma^0\psi(-x_i,t)$ and the corresponding conserved charge is
\begin{equation}\label{g1}
 Q=\int d^3x_iJ^{0}=i\int d^3x_i {\psi}^{\dagger}(x_i,t)\gamma^0\psi(-x_i,t).
\end{equation}
Hence $\delta\psi(x)$ corresponding to parity symmetry is obtained by comparing Eq.(\ref{g3}) and Eq.(\ref{g1}) as $\delta\psi(x)=\psi_p(x_i,t)=\gamma^0\psi(-x_i,t)$. Similarly for a time-reversal symmetry the conserved current is $J^{\mu}=i\bar{\psi}(x_i,t)\gamma^{\mu}\psi_T(x_i,t)$ and the corresponding conserved charge is
\begin{equation}\label{g4}
 Q=\int d^3x_iJ^{0}=-\int d^3x_i {\psi}^{\dagger}(x_i,t)\gamma^1\gamma^3\psi^*(x_i,-t).
\end{equation}
Therefore $\delta\psi(x)$ corresponding to time-reversal symmetry is obtained from Eq.(\ref{g3}) and Eq.(\ref{g4}) as $\delta\psi(x)=\psi_T(x_i,t)$. Following these, in the $\kappa$-deformed space-time, we get $\delta\hat{\psi}(x)=\hat{\psi}_P(x_i,t)$ corresponding to parity symmetry and $\delta\hat{\psi}(x)=\hat{\psi}_T(x_i,t)$ corresponding to time-reversal symmetry. Since the $\kappa$-deformed charges have the same form as that in the commutative case and $a$ dependent corrections come only from the deformed field and its adjoint, it is natural that parity and time-reversal operators, $\gamma^0$ and $i\gamma^1\gamma^3$, are same as that in the commutative space-time.
\section{Appendix B}
Here, we  discuss the construction of the generators of the space-time symmetry transformation in $\kappa$-deformed space-time. The expression for the $\kappa$-deformed conserved current becomes
\begin{equation}\label{e2}
\hat{J}^{\mu}=\bar{\hat{\psi}}(x)\hat{\Gamma}^{\mu}(\partial,-\overleftarrow{\partial})\delta\hat{\psi}(x)+h.c,
\end{equation}
where $\delta\hat{\psi}(x)=\hat{\psi}(x')-\hat{\psi}(x)$ is the infinitesimal change in the deformed Dirac field due to the infinitesimal change in the coordinates under the transformation, $x_{\mu}\to x_{\mu}'$. We determine this $\delta\hat{x}_{\mu}$ using explicit form of Dirac derivative in following equation as
\begin{equation}\label{f1}
 \delta\hat{x}_{0}=[D_0,\hat{x}_{0}]\theta_0,~~\delta\hat{x}_{i}=[D_i,\hat{x}_{i}]\theta_i
\end{equation}
Using Eq.(\ref{dr1}) in Eq.(\ref{f1}), we get $\delta\hat{x}_{\mu}$, valid up to first order in $a$, as
\begin{equation}\label{f2}
 \delta\hat{x}_{0}=\theta_0,~~\delta\hat{x}_{i}=\theta_i(1-ap_0)
\end{equation}
Thus we obtain $\delta\hat{\psi}(x)=-\theta_0\partial_0\hat{\psi}(x)+\theta_i(1-ap_0)\partial_i\hat{\psi}(x)$. Hence substituting this in Eq.(\ref{e2}), we obtain the deformed conserved current as
\begin{equation}\label{f3}
 \hat{J}^{\mu}=i\theta_{\nu}\bar{\hat{\psi}}\gamma^{\mu}\partial^{\nu}\hat{\psi}-iap_0\delta^{\nu i}\theta_{\nu}\bar{\hat{\psi}}\gamma^{\mu}\partial_i\hat{\psi}+\frac{a}{2}\theta_{\nu}\delta^{\mu i}\bar{\hat{\psi}}\gamma^0(\partial_i-\overleftarrow{\partial_i})\partial^{\nu}\hat{\psi}+h.c
\end{equation}
The general expression for the conserved current, obtained by varying the action, is given as
\begin{equation}\label{xy}
 J^{\mu}=-\frac{\partial \mathcal{L}}{\partial(\partial_{\mu}\phi)}\bar{\delta}\phi+T^{\mu\nu}\delta x_{\nu}, 
\end{equation}
where $\bar{\delta}\phi(x)=\phi'(x')-\phi(x)$. For a space-time translation, $\bar{\delta}\phi(x)=0$ and $\delta x_{\mu}=\theta_{\mu}$. Therefore we get the energy-momentum tensor by taking the derivative of ${J}^{\mu}$ with respect to $\delta x_{\nu}$, i.e., $\theta_{\mu}$. Hence we obtain the deformed energy-momentum tensor from Eq.(\ref{f3}) as $\hat{T}^{\mu\nu}=\frac{\partial \hat{J}^{\mu}}{\partial \theta_{\nu}}$. Thus 
\begin{equation}\label{f4}
 \hat{T}^{\mu\nu}=\frac{i}{2}\Big(\bar{\hat{\psi}}\gamma^{\mu}\partial^{\nu}\hat{\psi}+\bar{\hat{\psi}}\gamma^{\nu}\partial^{\mu}\hat{\psi}\Big)-i\frac{ap_0}{2}\Big(\bar{\hat{\psi}}\gamma^{\mu}\delta^{\nu i}\partial_i\hat{\psi}+\bar{\hat{\psi}}\gamma^{\nu}\delta^{\mu i}\partial_i\hat{\psi}\Big)
+\frac{a}{4}\Big(\delta^{\mu i}\bar{\hat{\psi}}\gamma^0(\partial_i-\overleftarrow{\partial_i})\partial^{\nu}\hat{\psi}+\delta^{\nu i}\bar{\hat{\psi}}\gamma^0(\partial_i-\overleftarrow{\partial_i})\partial^{\mu}\hat{\psi}\Big)+h.c
\end{equation}
Note that all $a$-dependent terms coming from $2^{nd}$ and $3^{rd}$ terms in the above vanish for $\hat{T}^{00}$.
 
Here we see that in the commutative limit, i.e., $a\to 0$, the energy momentum tensor of Dirac field reduces to the well-known form, $T^{\mu\nu}=\frac{i}{2}\Big(\bar{\hat{\psi}}\gamma^{\mu}\partial^{\nu}\hat{\psi}+\bar{\hat{\psi}}\gamma^{\nu}\partial^{\mu}\hat{\psi}\Big)+h.c.$ We use the $\hat{T}^{\mu\nu}$ calculated in Eq.(\ref{f4}) to obtain the deformed conserved momenta, $\hat{P}^{\mu}=\int d^3x \hat{T}^{\mu 0}$, as 
\begin{equation}\label{f5}
 \hat{P}^{\mu}=\int d^3x\bigg[\frac{i}{2}\Big(\bar{\hat{\psi}}\gamma^{\mu}\partial^{0}\hat{\psi}+\bar{\hat{\psi}}\gamma^{0}\partial^{\mu}\hat{\psi}\Big)-i\frac{ap_0}{2}\bar{\hat{\psi}}\gamma^{0}\delta^{\mu i}\partial_i\hat{\psi}
+\frac{a}{4}\delta^{\mu i}\bar{\hat{\psi}}\gamma^0(\partial_i-\overleftarrow{\partial_i})\partial^{0}\hat{\psi}+h.c\bigg]
\end{equation}
We also use deformed energy-momentum tensor expressed above, i.e., $\hat{T}^{\mu\nu}$, to construct the $\kappa$-deformed angular momentum operator from the relation $\hat{M}^{\mu\nu}=\int d^3x\big(\hat{T}^{\mu 0}\hat{x}^{\nu}-\hat{T}^{\nu 0}\hat{x}^{\mu}\big)=\int d^3x\big(\hat{P}^{\mu}\hat{x}^{\nu}-\hat{P}^{\nu}\hat{x}^{\mu}\big)$. Thus 
\begin{equation}\label{f6}
\begin{split}
 \hat{M}^{\mu\nu}&=\int d^3x\Bigg(\bigg[\frac{i}{2}\Big(\bar{\hat{\psi}}\gamma^{\mu}\partial^{0}\hat{\psi}+\bar{\hat{\psi}}\gamma^{0}\partial^{\mu}\hat{\psi}\Big)x^{\nu}+\frac{i}{2}ap_0\Big(\bar{\hat{\psi}}\gamma^{\mu}\partial^{0}\hat{\psi}+\bar{\hat{\psi}}\gamma^{0}\partial^{\mu}\hat{\psi}\Big)x^{\nu}-i\frac{ap_0}{2}\bar{\hat{\psi}}\gamma^{0}\delta^{\mu i}\partial_i\hat{\psi}x^{\nu}
\\&+\frac{a}{4}\delta^{\mu i}\bar{\hat{\psi}}\gamma^0(\partial_i-\overleftarrow{\partial_i})\partial^{0}\hat{\psi}x^{\nu}\bigg]
-\bigg[\frac{i}{2}\Big(\bar{\hat{\psi}}\gamma^{\nu}\partial^{0}\hat{\psi}+\bar{\hat{\psi}}\gamma^{0}\partial^{\nu}\hat{\psi}\Big)x^{\mu}+\frac{i}{2}ap_0\Big(\bar{\hat{\psi}}\gamma^{\nu}\partial^{0}\hat{\psi}+\bar{\hat{\psi}}\gamma^{0}\partial^{\nu}\hat{\psi}\Big)x^{\mu}\\&-i\frac{ap_0}{2}\bar{\hat{\psi}}\gamma^{0}\delta^{\nu i}\partial_i\hat{\psi}x^{\mu}
+\frac{a}{4}\delta^{\nu i}\bar{\hat{\psi}}\gamma^0(\partial_i-\overleftarrow{\partial_i})\partial^{0}\hat{\psi}x^{\mu}\bigg]+h.c\Bigg)
\end{split}
\end{equation} 
We see that in the commutative limit, i.e., $a\to 0$, the angular momentum tensor for the Dirac field becomes $M^{\mu\nu}=\int d^3x\Big(\frac{i}{2}\big(\bar{\hat{\psi}}\gamma^{\mu}\partial^{0}\hat{\psi}+\bar{\hat{\psi}}\gamma^{0}\partial^{\mu}\hat{\psi}\big)x^{\nu}
-\frac{i}{2}\big(\bar{\hat{\psi}}\gamma^{\nu}\partial^{0}\hat{\psi}+\bar{\hat{\psi}}\gamma^{0}\partial^{\nu}\hat{\psi}\big)x^{\mu}+h.c\Big)$. Straight forward calculations show that the $\hat{P}^{\mu}$ and $\hat{M}^{\mu\nu}$ derived in Eq.(\ref{f5}) and Eq.(\ref{f6}) satisfy the undeformed $\kappa$-Poincare algebra \cite{mel1}.

\subsection*{\bf Acknowledgments}
EH thanks SERB, Govt. of India, for support through EMR/2015/000622. VR thanks Govt. of India, for support through DST-INSPIRE/IF170622.

\end{document}